\documentstyle[12pt,epsfig,subfigure,amssymb,latexsym,aaspp4,natbib]{article}
\newcommand{\rb}[1]{\raisebox{1.5ex}[-1.5ex]{#1}}
\newcommand{\sqdeg}{\mbox{deg}^{2}}

\bibpunct{(}{)}{;}{a}{}{,}
\begin{document}
\begin{flushleft}
\raisebox{2cm}[-2cm]{\textsc{\small To appear in: Proceedings of Workshop on
    Subaru HDS, Tokyo, December 8.--10., 1999}}
\end{flushleft}
\lefthead{Christlieb \& Beers}
\righthead{Surveys for Metal-Poor Stars}
\title{Ongoing Large Surveys for\\
  Metal-Poor Stars in the Galactic Halo}
\author{Norbert Christlieb}
\affil{Hamburger Sternwarte, Germany\\
  E-mail: nchristlieb@hs.uni-hamburg.de}
\and
\author{Timothy~C. Beers}
\affil{Department of Physics \& Astronomy, Michigan State University,\\
  East Lansing, MI, USA\\
  E-mail: beers@pa.msu.edu}

\begin{abstract}
We report on two major surveys for metal-poor stars in the galactic halo, the
HK survey, and the Hamburg/ESO survey, which have been undertaken in order to
provide targets for high-resolution spectroscopy with the Subaru HDS and other
large telescopes.  We compare basic properties of these two surveys and their
current status, and add some historical remarks.

The candidate selection procedures of both surveys are described in detail.
We evaluate the candidate selection by comparing {\em effective yields\/}
(EYs) of the survey techniques for the identification of metal-poor stars. It
is found that EY for stars below [Fe/H]$=-2.0$ in the HES can be up to $\sim
80$\,\% for stars selected by automatic classification from machine-scanned
unwidened plates, whereas in the HK survey, where stars are selected by visual
inspection of widened survey plates, the EY is between 11\,\% and 32\,\%,
depending on whether a pre-selection based on $BV$ photometry has been
applied.

Finally, we describe techniques used for determining stellar parameters of the
survey stars by means of moderate resolution follow-up spectroscopy, and
additional $UBV$ photometry.  While follow-up observations of HES stars have
just been started, the HK survey has already produced a list of $\sim 4700$
stars with estimates of [Fe/H] typically precise to $\pm 0.2$\,dex, some 1000
of which have [Fe/H] $< -2.0$, and roughly 100 of which have [Fe/H] $< -3.0$.
\end{abstract}

\section{Introduction}

With the advent of several new 8\,m class telescopes, e.g. Subaru, and the VLT
telescopes, it is anticipated that many new insights into the nature of the
Galactic halo, the chemical evolution of our Galaxy, and the first stars to
have formed within it, will soon be in the offing.  However, it would be
impossible to obtain such exciting results if there where no large surveys
that can provide {\em targets\/} for high-resolution, high-$S/N$ observations
with these new instruments of discovery. In this article, we give a detailed
comparison of two such surveys, namely the HK survey \citep{BPSI,BPSII}, and
the Hamburg/ESO survey \citep[HES; ][]{HESpaperI,HESpaperIII}. For a
comprehensive review of past, present and future surveys for metal-poor stars
we refer the reader to \cite{PopIIIbyDemand}.

\section{Basic properties of the surveys}

In this section we compare basic properties of the HK and HE surveys, 
and add some historical remarks. An overview of the survey properties is given
in Tab. \ref{HKHESproperties}.

\begin{table}[htbp]
  \begin{center}
    \begin{tabular}{llll}\tableline\tableline
         & & HK survey & \rule{0ex}{2.3ex}HES \\\tableline
   & north & \rule{0ex}{2.3ex}0.6\,m Burrell Schmidt & --- \\[-0.3ex]
   \rb{Telescope} & south & 0.6\,m Curtis Schmidt & 1\,m ESO Schmidt\\[0.5ex]
   \multicolumn{2}{l}{Magnitude range} & $11.0\lesssim B\lesssim 15.5$ &
                            $14.0\lesssim B\lesssim 17$ \\[0.5ex]
   \multicolumn{2}{l}{Widened?} & yes & no \\[0.5ex]
   & north & $2800\,\Box^{\circ}$ & --- \\[-0.3ex]
   \rb{Area} & south & $4100\,\Box^{\circ}$ & $7600\,\Box^{\circ}$\\[0.5ex]
   \multicolumn{2}{l}{Objective prism} & $4^{\circ}$ & $4^{\circ}$\\[0.5ex]
   \multicolumn{2}{l}{Dispersion} & $180$\,{\AA}/mm & $450$\,{\AA}/mm\\[0.5ex]
   \multicolumn{2}{l}{Spectral resolution}
   & $\sim 5\,${\AA} &  $\sim 10\,${\AA} at \ion{Ca}{2}~K\\[0.5ex]
   \multicolumn{2}{l}{Photographic emulsion} & 103a-O/IIa-O & IIIa-J \\[0.5ex]
   \multicolumn{2}{l}{Filter?} & interference/Ca~H+K & no \\[0.5ex]
   \multicolumn{2}{l}{Wavelength range}
   & $3875\,${\AA}$<\lambda < 4025\,${\AA} & $3200\,${\AA}$<\lambda<5200\,${\AA}\\[0.5ex]
   \multicolumn{2}{l}{Candidate selection} & visual inspection & automated
   \\\tableline\tableline
 \end{tabular}
\end{center}
\caption{\label{HKHESproperties} Comparison of the HK survey and the
  Hamburg/ESO survey (HES).}
\end{table}

\subsection{HK survey}

In 1978, G. Preston and S. Shectman of the Carnegie Observatories of
Washington started an objective-prism survey for the discovery of numerous
metal-poor and field horizontal-branch stars in the Galaxy.  This was at a time
when it was generally assumed that stars more metal-deficient than the most
metal-poor globular clusters (${\rm [Fe/H]} \sim -2.5$) do not exist. In 1983
Beers joined the team, and later expanded the survey with an additional 240
plates in the southern and northern hemispheres.  This survey, once referred
to as the ``Preston-Shectman Survey,'' is now widely known as the ``HK
survey.''  This is because in addition to a 4$\arcdeg$ objective prism (leading
to a seeing-limited spectral resolution of $\sim 5$\,{\AA}), an interference
filter was mounted on the plate holder to limit the wavelength coverage to
$\sim 150\,${\AA} centered on the \ion{Ca}{2}~H+K resonance lines, effectively
reducing the sky background level so that long exposures (typically 90 minutes)
could be obtained.  By 1992, 308 acceptable-quality plates were obtained (275
of which are unique) with the 60\,cm Burrell Schmidt (northern hemisphere) and
Curtis Schmidt (southern hemisphere) telescopes, each plate covering
$5\arcdeg\times5\arcdeg$ of the sky.  Further extension of the survey area was
prevented by the shortage of photographic plates with 103a-O and IIa-O
emulsions.

\subsection{HES}\label{HESproperties}

\begin{figure}[htbp]
  \begin{center}
    \leavevmode
    \raisebox{5mm}[5mm]{\epsfig{file=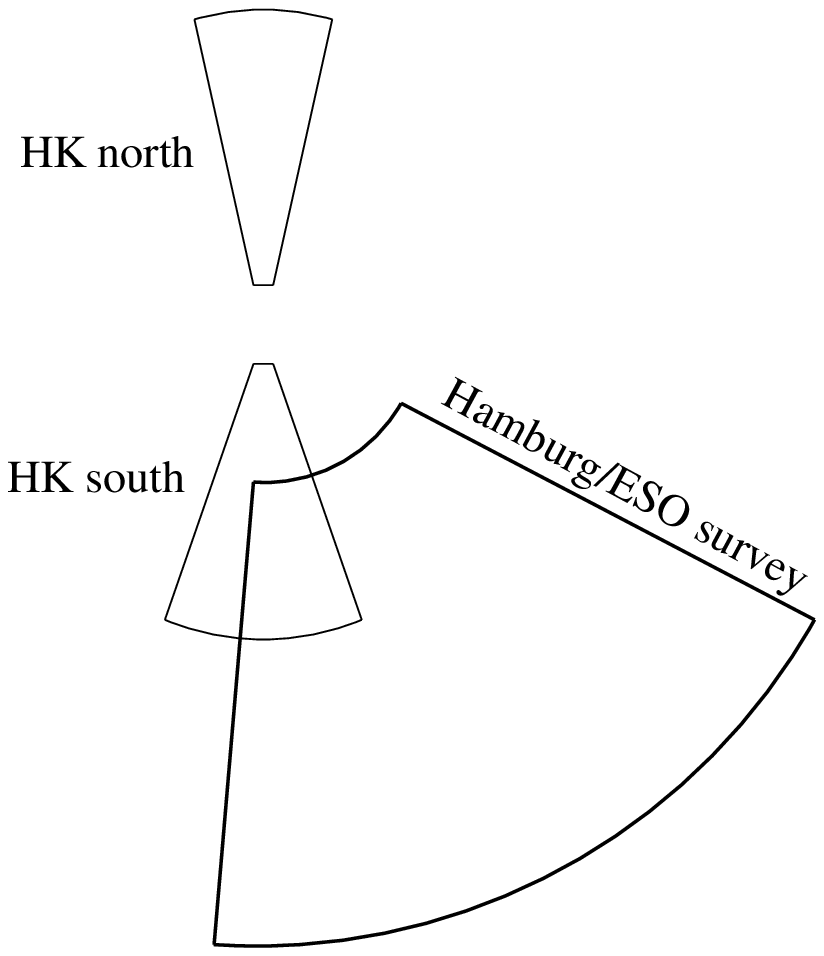, height=6.0cm}}
    \hspace{5mm}
    \epsfig{file=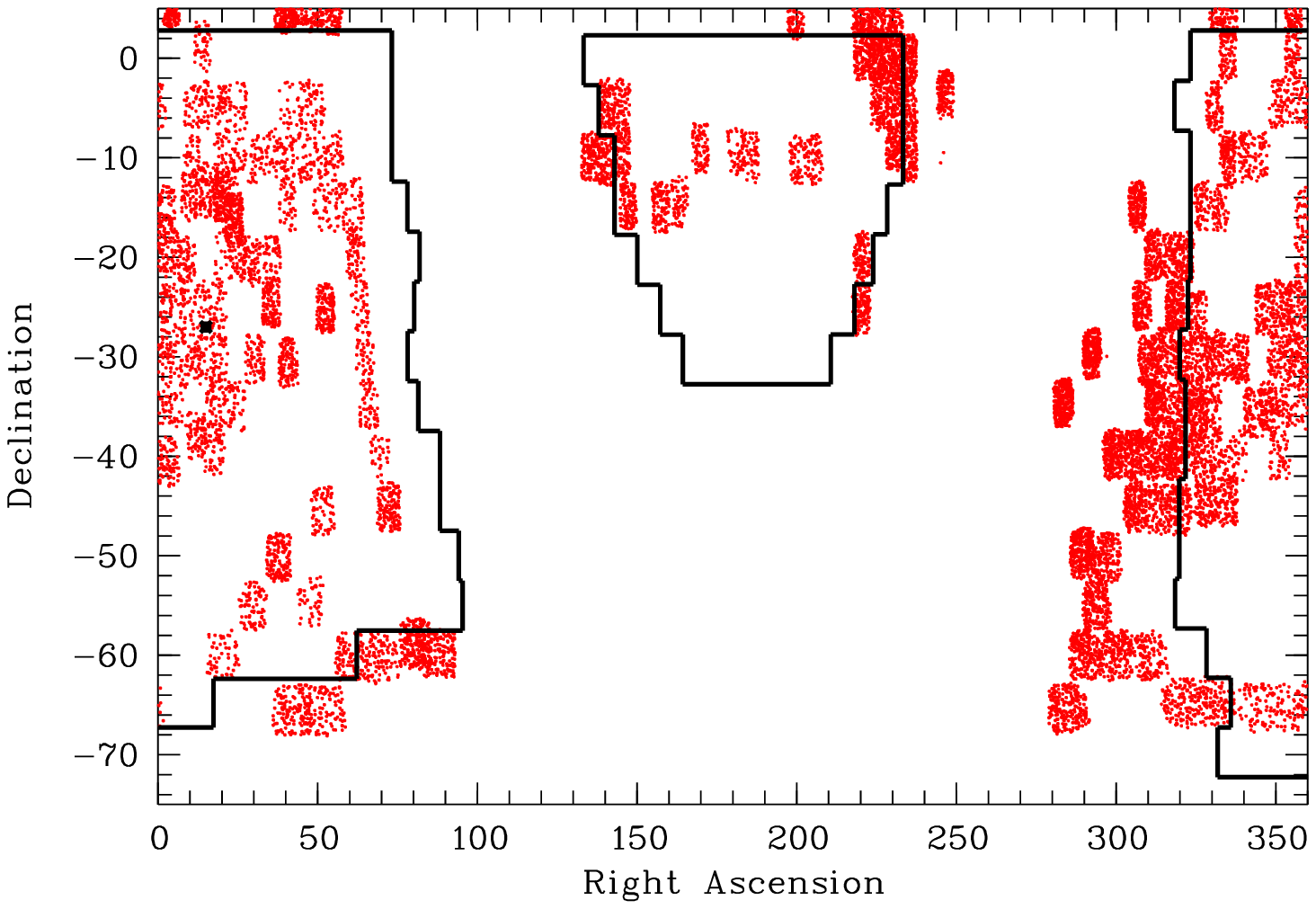, clip=, width=10cm,
      bbllx=49, bblly=423, bburx=471, bbury=712}
  \end{center}
  \caption{\label{HESHKareavolcomp} Left panel: Comparison of HES and HK
    survey volumes; right panel: Comparison of HES area (framed) with HK
    survey area. Dots denote all HK survey candidates in the southern
    hemisphere.}
\end{figure}

The HES was started in 1989 as an ESO Key Programme (P.I.: D. Reimers; Project
Manager: L. Wisotzki). Its main aim is to find bright quasars. However, it was
recognized right at the start that with the HES' seeing-limited spectral resolution
of $\sim 15$\,{\AA} at H$\gamma$, it would be feasible to do a lot of
interesting {\em stellar\/} work as well. In 1994, Christlieb joined the HES
group, and began development of methods for the systematic exploitation of the
stellar content of the HES.  By that time almost half of the objective-prism
plates had already been taken in service mode, with the ESO 1\,m-Schmidt
telescope and its $4\arcdeg$ prism, and most of the data reduction software had
been developed by L. Wisotzki and T. K\"ohler.  Because the HES plates were
taken {\em without\/} a filter, resulting in a wavelength coverage of
$3200\,\mbox{{\AA}}<\lambda<5200\,\mbox{{\AA}}$, some care had to be taken with
the identification of overlapping spectra.  Moreover, since the
HES was primarily a quasar survey, it was deemed not useful to work in fields
with too high foreground extinction.  As a result, the main criteria defining
the HES survey area are the mean star density, $\rho$, and column density of
neutral hydrogen, N$_{\mbox{\scriptsize H}}$:
\begin{eqnarray*}
        \rho &<& 100\mbox{ stars}/\sqdeg\\
        \mbox{N}_{\mbox{\scriptsize H}} &<& 10^{21}\,\mbox{cm}^{-2}.
\end{eqnarray*}
This roughly corresponds to $|b|\gtrsim 30\arcdeg$. The survey is restricted
to the southern hemisphere, i.e. $\delta<+2.5\arcdeg$, but $\sim 50\,\%$ of
the HES fields are located at $\delta>-25\arcdeg$, so that half of the stars
found in the HES are easily reachable for Subaru. That is, on Mauna Kea they
are at $\sec z<2.0$ for several hours per night in the appropriate months.
HES areas in common with the HK survey are shown in Fig. \ref{HESHKareavolcomp}.

\begin{figure}[htbp]
  \begin{center}
    \leavevmode
    \epsfig{file=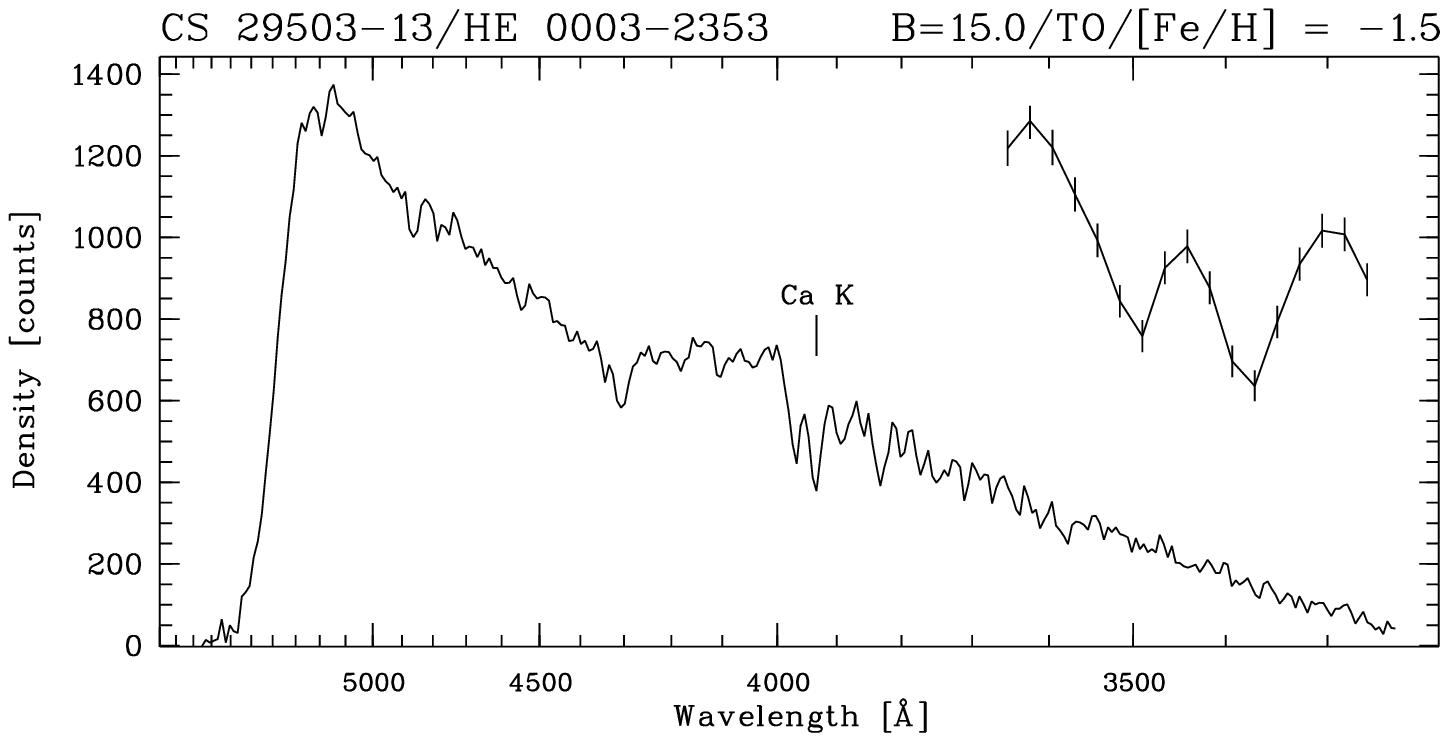, clip=, width=12cm,
      bbllx=53, bblly=449, bburx=470, bbury=638}\\[1mm]
    \epsfig{file=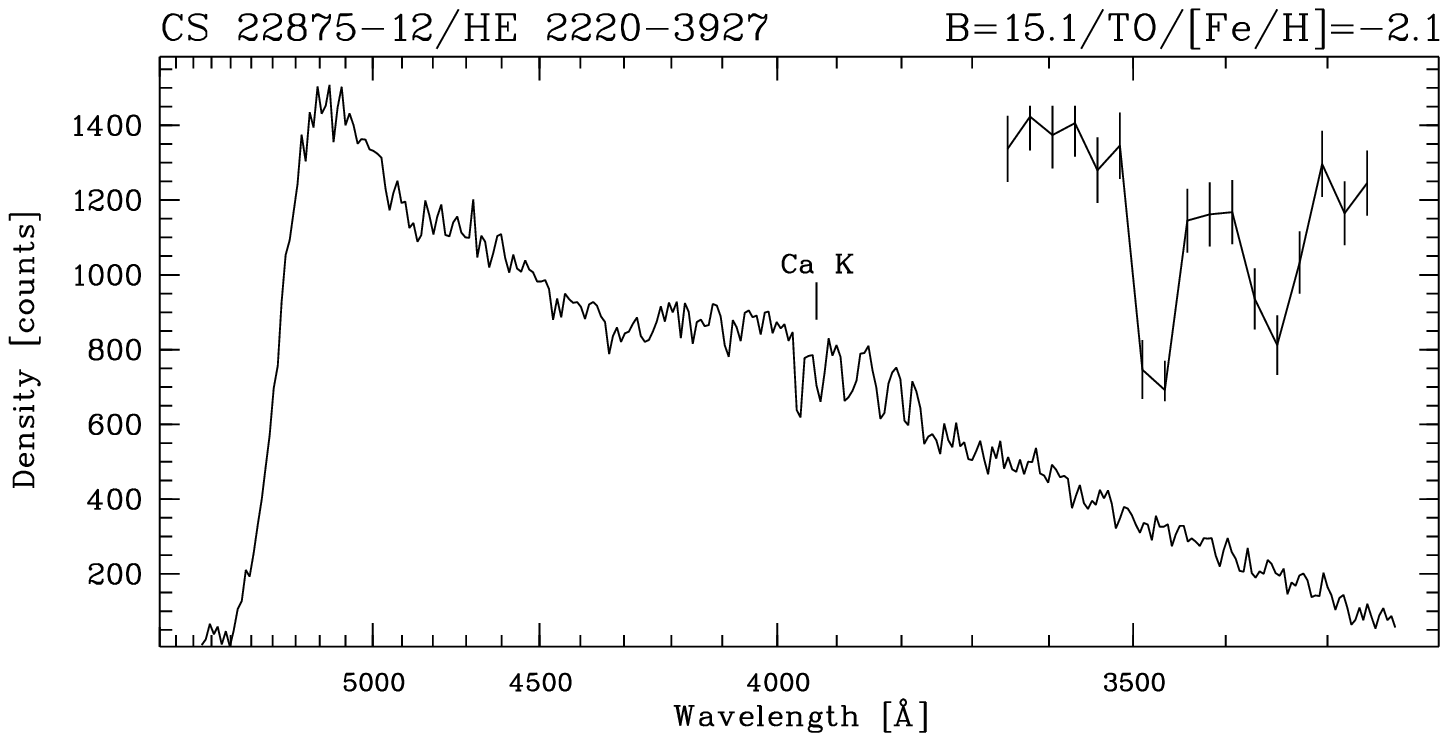, clip=, width=12cm,
      bbllx=53, bblly=449, bburx=470, bbury=638}\\[1mm]
    \epsfig{file=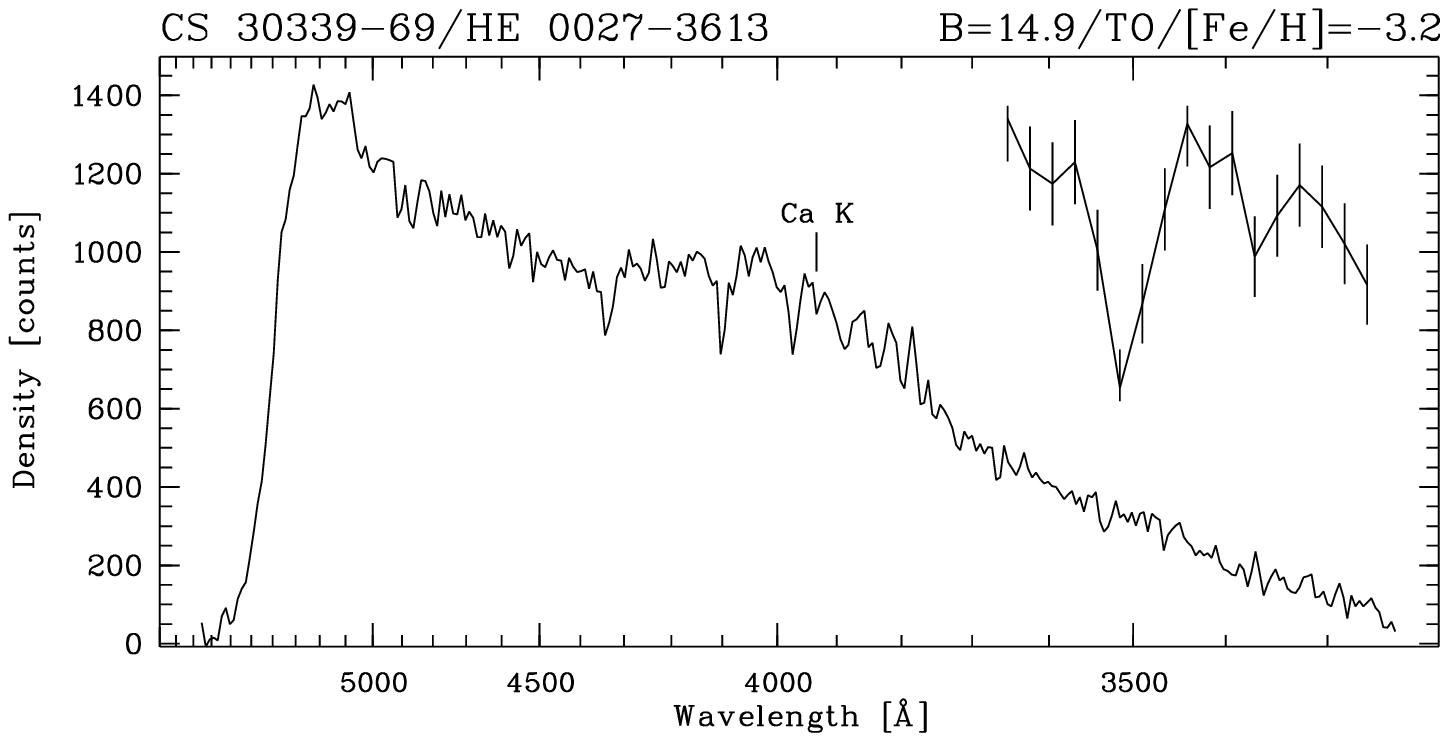, clip=, width=12cm,
      bbllx=53, bblly=426, bburx=470, bbury=638}
    \caption{\label{HESmphsexamples} Examples of HES spectra of metal-poor
      turnoff stars discovered in the HK survey. In the upper right corner of each
      plot a blow-up of the \ion{Ca}{2}~H+K region is shown, with an overplot
      of the pixelwise 1\,$\sigma$ noise. Note that the detection of
      \ion{Ca}{2}~K in the lower spectrum is not significant. Metal abundances
      are on the re-calibrated HK survey scale of \cite{HKrecalib}.}
  \end{center}
\end{figure}

By October 1998, just before de-comissioning of the ESO Schmidt telescope, the
last HES plate was taken. Today, all 383 plates defining the survey have been
scanned in Hamburg using a PDS~1010G microdensitometer. The HES database now
consists of $\sim 4,000,000$ digital, extracted, wavelength calibrated,
non-overlapping spectra with mean $S/N>5$ (for example spectra see Fig.
\ref{HESmphsexamples}). Note that the elimination of overlapping spectra reduces
the survey area from a nominal $9575\,\sqdeg$ to an effective area of $\sim
7600\,\sqdeg$, similar to the total area of the HK survey, where overlapping
spectra were not such a severe problem.

The use of a larger telescope, and a $2$ times lower resolution of the HES
compared to the HK survey, results in a limiting magnitude of about $B=17.5$.
However, we restricted the selection of metal-poor candidate stars in the HES
to $S/N>10$, because it was found that below this $S/N$ level it is extremely
difficult to select objects by the absence of individual spectral lines (e.g.,
the \ion{Ca}{2}~K line in case of metal-poor stars). In result, the faintest
low-metallicity candidates in the HES sample ``only'' reach $B=17$, about
$1.5$ magnitudes deeper than the HK survey. Spectra of bright objects close to
saturation where excluded from the search for metal-poor stars, too, because
at high illumination, when the characteristic curve of the photographic
emulsion gets flatter (at the ``shoulder''), the contrast between continuum
and spectral lines gets weaker, and apparently {\em all\/} stars have weak
lines. The saturation threshold choosen in the HES corresponds to $B\sim
14.0$. Taking the common area of both surveys and their magnitude ranges into
account, the HES can increase the total survey volume for metal-poor stars by
a factor of 8 compared to the HK survey alone (see also Fig.
\ref{HESHKareavolcomp})!

\section{Candidate selection}

\subsection{HK survey}

\begin{figure}[htbp]
  \begin{center}
    \leavevmode
    \epsfig{file=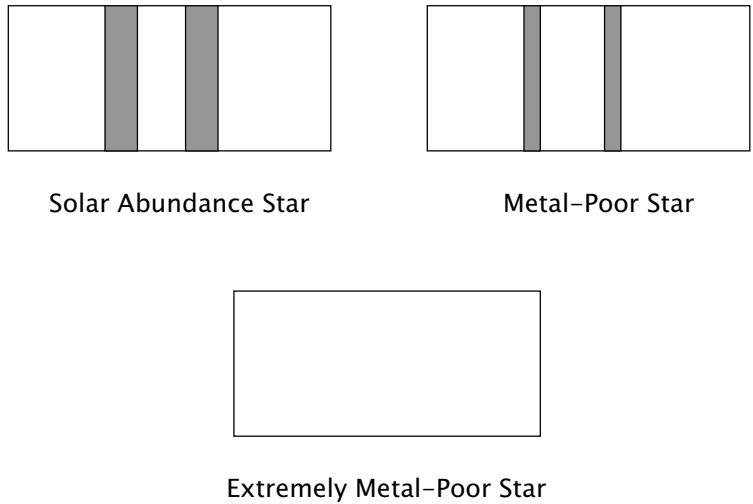, clip=, width=10cm,
      bbllx=72, bblly=422, bburx=492, bbury=705}
  \end{center}
  \caption{\label{HKspectra} Drawings of HK survey spectra.}  
\end{figure}

Candidate selection in the HK survey was done by visual inspection of the
widened objective-prism spectra with a binocular $10\times$ microscope. Each
plate was inspected twice, with a lag time of a month or more between the two
inspections.  Candidates were identified on the basis of the observed
strengths of their \ion{Ca}{2} lines (see Fig. \ref{HKspectra}), and grouped
into rough categories based on this criteria (e.g., possibly metal-poor,
metal-poor, and extremely metal-poor).  Positions of the candidates were noted
on the plates, and coordinates for each candidate were measured later
(individually, with Grant machines).  In this process, a total of about
$10\,000$ metal-poor candidates was selected (roughly half of which have had
medium-resolution follow-up spectroscopy obtained to date).

Note that since the visual inspection process was made in the absence of any
information about the stellar colors (hence temperatures), it was expected that
the HK survey candidates would carry a rather severe temperature-related bias,
in the sense that cooler metal-deficient stars would likely be missed because
of the apparent strength of their \ion{Ca}{2} lines at lower temperatures. In
addition, stars of high temperature with intermediate abundances would be
included in greater number than might be desired because of the apparent
weakness of their \ion{Ca}{2} lines.  These biases become less of a problem at the
lowest metallicities, below [Fe/H]$= -2.0$, where the \ion{Ca}{2} lines of even quite
cool stars are difficult to detect at the resolution of the HK survey.

\subsection{HES}

For the present, we have restricted the selection of metal-poor stars in the
HES to the color range $0.3<\bv<0.5$, because we decided to focus at first on
main-sequence turnoff stars.  One of the most interesting applications for
these stars is individual age determination based on precise stellar parameters
obtained spectroscopically from high-resolution, high $S/N$ observations.
However, with a few adaptions the selection procedures described below can
easily be used for cooler stars, too, and they {\em will\/} be used for that in
the near future, provided that financial support for the continuation of
this project is obtained.

Candidate selection in the HES is done by two techniques: The \ion{Ca}{2}~K
index method, and via automatic classification. In the former, stars are
selected when their \ion{Ca}{2}~K line is significantly weaker than
``normal.''  What is ``normal'' is determined by a least squares fit of a 2nd
order polynomial to the \ion{Ca}{2}~K index relative to the parameter
\verb|x_hpp2|, which is the half power point of the density distribution of
the objective prism spectra in the wavelength range
$3890\,\mbox{\AA}<\lambda<5360\,\mbox{\AA}$. \verb|x_hpp2| is well-correlated
with $\bv$ color, with a $1\,\sigma$ dispersion of $\sim 0.1$\,mag. Spectra
having a \ion{Ca}{2}~K index which is more than $3\,\sigma$ below the
polynomial fit are selected as metal-poor candidates.

Below we give an outline of metal-poor star selection by automatic
classification in the HES. A more detailed description of the method, and all
procedures involved (e.g. conversion of flux spectra to artificial
objective-prism spectra, automatic feature detection, etc.) will be given in
an upcoming paper (Christlieb et al. 2000, in preparation).

For automatic classification we use a learning sample consisting of 45
classes defined by the following grid points:
\begin{eqnarray*}
  T_{\mbox{\scriptsize eff}} &=& 5800\,\mbox{K},\, 6400\,\mbox{K},\,
  6800\,\mbox{K}\\
  \log g &=& 2.2,\, 3.8,\, 4.6\\
  \left[\mbox{Fe/H}\right] &=& -0.9,\,-1.5,\,-2.1,\,-2.7,\,-3.3\\ 
\end{eqnarray*}
The learning sample has been constructed by converting model spectra to
simulated objective-prism spectra by reduction of the resolution, and
convolution with the effective spectral response of the photographic emulsion
and the transmission function of the prism. The model spectra have been kindly
provided by J. Reetz and T.  Gehren (Universit\"ats-Sternwarte M\"unchen,
Germany).


Nine automatically detected features are used for classification.  These are
the strengths of \ion{Ca}{2}~K, measured by an absorption line fit, and by an index
method; the sum of the equivalent widths of H$\beta$, H$\gamma$ and H$\delta$;
the half-power point \verb|x_hpp2|; the three principal components of metal-poor
star spectra that account for $90$\,\% of the variance in the learning sample;
and the Str\"omgren coefficient $c_1$, which is directly determined from the
objective-prism spectra by integration over the relevant wavelength range.
The $c_1$ index has been calibrated against HK survey metal-poor stars of
\cite{Schusteretal:1996} present on HES plates. The accuracy achieved in this
effort is $\sigma_{c_1}=0.05$\,mag.

The values of the above quantities are organized into ``feature
vectors'' $\vec{x}$. Class-conditional probabilities $p(\vec{x}|\Omega_i)$ of
the learning sample are modelled by multivariate normal distributions, i.e.,
\begin{equation}\label{multigauss}
   p(\vec{x}|\Omega_i) = \frac{1}{(2\pi)^{d/2}\sqrt{|\Sigma_i|}}
   \exp\left\{-\frac{1}{2}\left(\vec{x}-\vec{\mu}_i\right)
   \Sigma^{-1}_{i}\left(\vec{x}-\vec{\mu}_i\right)'\right\},
\end{equation}
where $i$ denotes class number, $\vec{\mu}_i$ the mean feature vector of class
$\Omega_i$, and $\Sigma_i$ the covariance matrix of class $\Omega_i$. Using
Bayes' theorem,
\begin{displaymath}
  p(\Omega_i|\vec{x})=\frac{p(\Omega_i)p(\vec{x}|\Omega_i)}
  {\sum\limits_{\forall i}p(\Omega_i)p(\vec{x}|\Omega_i)},
\end{displaymath}
posterior probabilities $p(\Omega_i|\vec{x})$ can then be calculated. We
assume equal prior probabilities $p(\Omega_i)$ for all classes present in the
learning sample.

A spectrum of unknown class, with given feature vector $\vec{x}$, is classified
according to Bayes' rule: {\em Assign the spectrum to the class with the
highest posterior probability $p(\Omega_i|\vec{x})$}. This rule minimizes the
total number of misclassifications if the {\em real\/} distribution of
class-conditional probabilities $p(\vec{x}|\Omega_i)$ is used. It remains to
be tested {\em quantitatively\/} if the class-conditional probabilities follow
multivariate normal distributions; however, this has been tested {\em
qualitatively\/} by visual inspection of the distributions at the computer
screen.

Non-mathematically speaking, Bayes' rule assigns the class with the
highest {\em relative\/} resemblance to each spectrum to be classified.
However, it is ignorant of the {\em absolute\/} resemblance: A spectrum with
feature vector $\vec{x}$ may be assigned to a class with {\em very low\/}
posterior probability $p(\Omega_i|\vec{x})$, if $p(\Omega_i|\vec{x})$ is even
lower for all other classes. This means that a class is assigned to {\em
  all\/} spectra, even to ``garbage spectra'' which have been disturbed, for
instance, by plate artefacts. Therefore, it is useful to apply the following
rejection rule: {\em Reject an object from classification to class
  $\Omega_i$, if $a.i.(\Omega_i;\vec{x}) > \beta\,$}. The parameter $\beta$ is
a threshold to be chosen, and the parameter $a.i.$ is the {\em
  atypicality index\/} suggested by \cite{Aitchisonetal:1977},
\begin{displaymath}
  a.i.(\Omega_i,\vec{x}) = \Gamma\left\{\frac{d}{2};\frac{1}{2}
      \left(\vec{x}-\vec{\mu}_i\right)
      \Sigma^{-1}_{i}\left(\vec{x}-\vec{\mu}_i\right)'\right\},
\end{displaymath}
where $\Gamma(a;x)$ is the incomplete gamma function and $d$ the number of
features used for classification. Use of the above rejection criterion is
identical to performing a $\chi^2$ test of the null hypothesis $H_0$ that an
object with feature vector $\vec{x}$ belongs to class $\Omega_i$ at
significance level $1-\beta$, against the alternative hypothesis $H_1$ that it
{\em does\/} belong to class $\Omega_i$. We reject the null hypotheses, if its
significance level is {\em low}, i.e., if it is very {\em un\/}likely that a
feature vector $\vec{x}$ is observed for class $\Omega_i$, given the
multivariate normal distributions (\ref{multigauss}) are the {\em real\/}
distributions of the class-conditional probabilities $p(\vec{x}|\Omega_i)$.

Note that the automatic classification programs are fed only with a subset
of all spectra present on each HES plate. As already mentioned in section
\ref{HESproperties}, only spectra with $S/N>10$ and $B\gtrsim 14$ are
considered. Moreover, spectra outside of the range $0.3<\bv<0.5$ are excluded,
where $\bv$ is known to $\pm 0.1$\,mag from the calibration of \verb|x_hpp2|.

Below we summarize the selection criteria for metal-poor stars in the HES for
the selection by automatic classification. Pre-selection of spectra to which
automatic classification procedures are applied is done by criteria (1)--(3);
(4) and (5) use the results of automatic classification, and (6) is a
rejection criterion corresponding to a $\chi^2$ test at a $3\,\sigma$ level.
\begin{enumerate}
  \item[(1)] $0.3<\bv<0.5$
  \item[(2)] $(S/N)_{\mbox{\scriptsize HES}}>10\Longleftrightarrow B\lesssim 17.0$
  \item[(3)] Photographic density $D$ below saturation threshold 
  \item[(4)] $\log g\ge3.8$
  \item[(5)] $\left[\mbox{Fe/H}\right]\le-2.7$
  \item[(6)] $a.i.<0.99$.
\end{enumerate}

The final step of the selection is visual inspection of the automatically
selected spectra at the computer screen. This step is done for {\em both\/}
selections described above. Visual inspection is necessary for identification
of plate artefacts (e.g. scratches or emulsion flaws), and for rejection of
obviously misclassified spectra, i.e. spectra which clearly show a strong
\ion{Ca}{2}~K line. The remaining candidates are divided into three classes
according to the appearance of the \ion{Ca}{2}~K line region: ``class a''
candidates show clearly no line; in spectra of ``class b'' candidates it is
unclear if they have a line, and ``class c'' candidates {\em do\/} show a
\ion{Ca}{2}~K line, but however, a weak one. Typically, only 10\,\% of the
candidates belong to class a or b, 40\,\% belong to class c, 25\,\% are
misclassifications, and further 25\,\% are disturbed spectra.

\section{Effective yields}

As has been pointed out by \cite{PopIIIbyDemand}, the {\em effective yield\/}
(EY) of a detection method is one of the most important properties of a survey
for metal-poor stars. EY is defined as follows:
\begin{displaymath}
  \mbox{EY}_{x}:=\frac{N_{\mbox{\footnotesize
        stars}}\,\,\mbox{with}\,\,\left[\mbox{Fe/H}\right]\,<\,x}{
    N_{\mbox{\footnotesize stars, observed}}}.
\end{displaymath}

When EYs for different surveys are compared, it is crucial to make sure that
the comparison is done on the same abundance scale. In case of the HK survey
and the HES, it was found that metallicities derived from the first-pass
analysis of the HES follow-up spectroscopy are $\sim 0.5$\,dex {\em higher\/}
on average, than obtained from the \cite{HKrecalib} re-calibration.  That
is,
\begin{displaymath}
  \left[\mbox{Fe/H}\right]_{\mbox{\scriptsize HK}} =
  \left[\mbox{Fe/H}\right]_{\mbox{\scriptsize HES}} - 0.5.
\end{displaymath}
This offset of the scales is primarily due to the different temperature scales
adopted in the two methods. In the HK survey, effective temperatures are
(implicitly) derived from $BV$ photometry, whereas in the HES, Balmer lines are
used. The abundance scale previously employed in the HK survey, e.g. in
\cite{BPSII}, is known to be an {\em additional\/} $0.2$\,dex lower for the
lowest metallicity stars \citep[see][]{Beersetal:2000}.

\begin{figure}[htbp]
  \begin{center}
    \leavevmode
    \epsfig{file=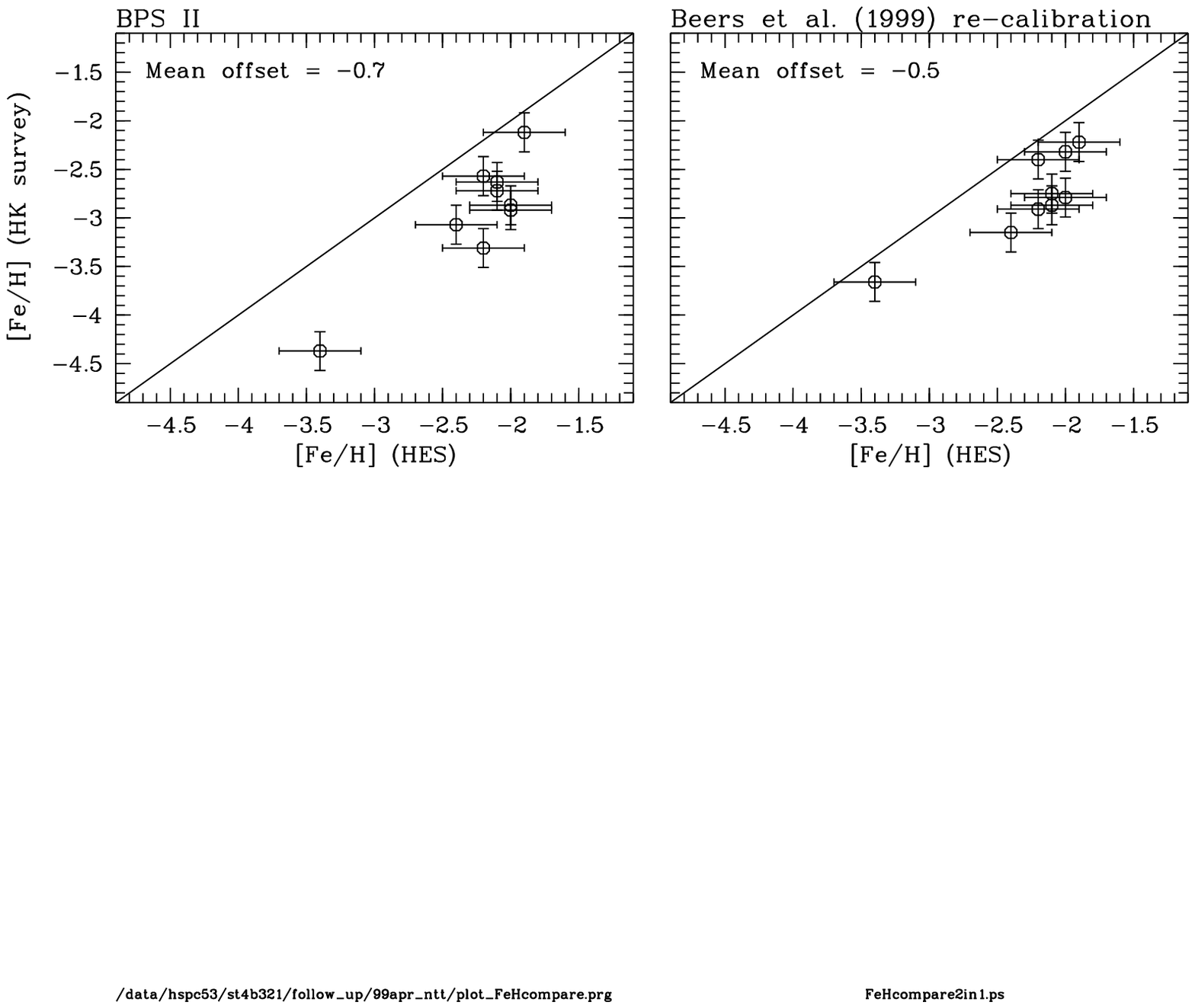, clip=, width=14.5cm, bbllx=68, bblly=285,
      bburx=526, bbury=466}
  \end{center}
  \caption{\label{FeHcompare} Comparison of HES and HK survey abundance
    scales. Error bars are $0.2$\,dex for the HK survey values, and $0.3$\,dex
    for HES values.}
\end{figure}

Thus far, only nine stars have been analyzed with {\em both\/} follow-up
techniques (see Fig. \ref{FeHcompare}), and there is especially a paucity of
comparison objects at $\left[\mbox{Fe/H}\right]_{\mbox{\scriptsize HES}}<-2.5$.
However, the derived trend is consistent for all data points. Only turnoff
stars have been used in the comparison; therefore, it can not be excluded that
the abundance difference is less (or even more) pronounced for cooler stars.

For this discussion, we restrict our EY comparison to turnoff stars in the
color range $0.3<\bv<0.5$, and carry out the comparison after an offset of
$0.5$\,dex has been subtracted from the HES metallicities.

In order to explore what the {\em highest possible\/} EY in the HES is, we
observed a sample of 58 HES metal-poor candidates with EMMI at the ESO NTT.
The stars have been selected by automatic classification, and have been
assigned to candidate classes a or b in the visual inspection. EY of stars at
[Fe/H] $<-2.0$ for this sample is 80\,\% (see Tab. \ref{EYcompare}). This has
to be compared with 11\,\% or 32\,\% in the HK survey, depending on whether a
pre-selection based on $B-V$ color has been made or not, respectively. We have
already obtained data for evaluation of the \ion{Ca}{2}~K index selection technique in
the HES, so that EY of that technique will be known soon.

\begin{table}
  \begin{center}
    \begin{tabular}{lcc}\tableline\tableline
      \rule{0ex}{2.3ex} Survey/selection method &
      $\mbox{EY}_{-2.0}$ & $\mbox{EY}_{-2.5}$\\\tableline
      HK survey/without $B-V$ pre-selection\rule{0ex}{2.3ex} & 11\,\% & 4\,\%\\
      HK survey/with $B-V$ pre-selection\rule{0ex}{2.3ex} & 32\,\% & 11\,\%\\
      HES/automatic classification & 80\,\% & 27\,\%\\\tableline\tableline
    \end{tabular}
  \end{center}
  \caption{\label{EYcompare} Comparison of effective yields (EY) of metal-poor
    turnoff stars of the HK survey and the HES. [Fe/H] is on the re-calibrated
    HK survey scale of \cite{HKrecalib}.}
\end{table}

\begin{figure}[htbp]
  \begin{center}
    \leavevmode
    \epsfig{file=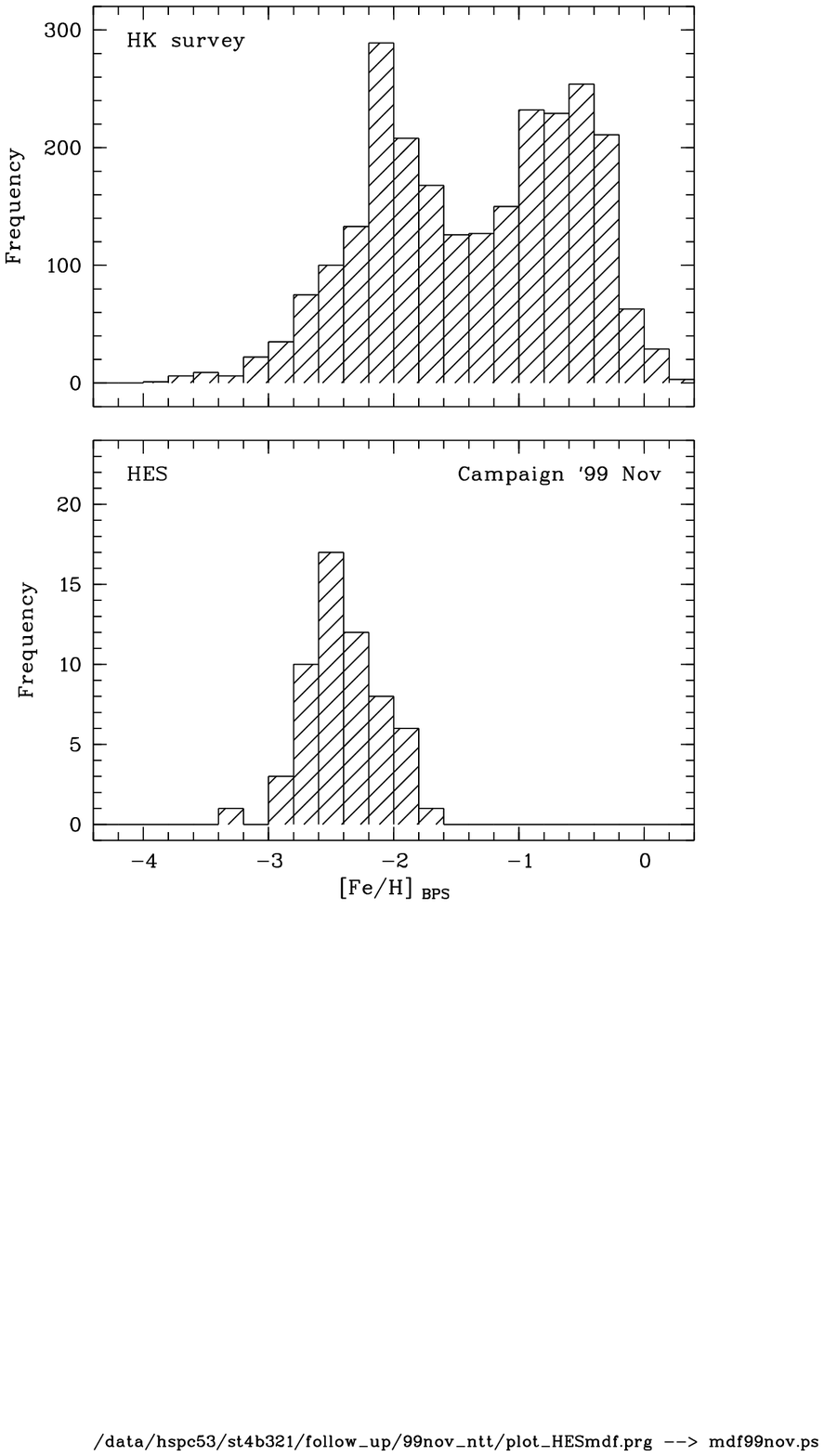, clip=, bbllx=106, bblly=327, bburx=399,
      bbury=712, width=10cm}
  \end{center}
  \caption{\label{MDFcompare} Metallicity distribution function  of stars
    in the color range $0.3<\bv<0.5$ from the HK survey (upper panel), and of
    a sample of 58 stars selected by automatic classification in the HES.}  
\end{figure}

\section{Follow-up techniques}

Because of the low quality of objective-prism spectra, and because of their
limited spectral resolution, prism surveys for metal-poor stars can only
provide, in general, {\em candidate\/} identifications.  Note that experiments
being conducted by J. Rhee, as part of his thesis work at Michigan State,
based on neural-network analysis of line strengths for \ion{Ca}{2}~H and K
obtained {\em directly\/} from automated scans of the HK survey plates, have
shown that it might be possible to assign metallicity estimates from the prism
spectra themselves, at least in a statistical sense \citep{Rheeetal:1999}.
However, in most applications to date, estimates of [Fe/H] and other stellar
parameters have to be derived by means of spectroscopic (and, for some
techniques, also photometric) follow-up observations. This intermediate step
has to be done with some care, because one doesn't want to spend significant
amounts of large telescope time for obtaining high-resolution, high-$S/N$
spectra of ``garden variety'' stars having as much as $\slantfrac{1}{{10}}$ of
the solar metal abundance!

\subsection{The \ion{Ca}{2}~K-index and ACF Methods}

For candidate low-metallicity stars in the HK survey, medium resolution
(1--2\,{\AA}) spectroscopy and broadband $BV$ photometry are used to obtain
metallicity estimates using two separate techniques.  The first technique
relies on the assumption that the strength of the \ion{Ca}{2}~K line tracks the
overall stellar [Fe/H], an assumption which is particularly good for stars with
[Fe/H] $\le -1.5$. The second is based on an Auto-Correlation Function
\citep[ACF, originally described by][]{Ratnatunga/Freeman:1989} of a stellar
spectrum.  The ACF method is particularly good for stars with [Fe/H] $> -1.5$,
where the \ion{Ca}{2}~K line begins to saturate with increasing metal abundance.
\cite{HKrecalib} discuss this calibration, and demonstrate, based on
comparisons with some 550 stars with external high-resolution abundance
estimates, that these approaches used in combination yield abundance
determinations with small scatter (on the order of 0.15--0.20\,dex) over the
entire range of stellar abundances we expect to find in the Galaxy ($-4.0 \le
{\rm[Fe/H]} \le 0.0$).

In a large collaborative effort involving many astronomers from the U.S.,
Europe, and Australia, $\sim 4700$ HK survey metal-poor candidates have had
spectroscopy obtained, and roughly half of them now have available $BV$
photometry.

\subsection{The ``all in one shot''-technique}

Due to limited telescope time available for follow-up observations, it would
be desirable to obtain estimates of stellar parameters, e.g., [Fe/H],
$T_{\mbox{\scriptsize eff}}$, and $\log g$, purely spectroscopically, {\em
  without\/} the need for additional photometry.  The first approach attempted
with the HES follow-up made use of comparisons with synthetic spectra.
However, it turned out that the choice to employ the \ion{Mg}{1}~b lines as
gravity indicators led to a number of difficulties.  For example, satisfactory
results required high $S/N$ ($>50$) spectra, which are very time consuming to
obtain for the fainter stars. Furthermore, at [Fe/H]$\lesssim -2.5$ and
turnoff temperatures, \ion{Mg}{1}~b is so weak that it is not sensitive to gravity
anymore.  Finally, the comparison of follow-up spectra with synthetic spectra
has to be done manually at the computer screen, which is a time sink as well.

As an alternative, the ``all in one shot''-technique described below was
developed.  It is fast, since for each star a single spectrum with $S/N\sim
30$ at \ion{Ca}{2}~K is all that is required, and data analysis can be done
fully automatically.

Spectrophotometry of each candidate is obtained with a wide slit ($\gtrsim
3\times$ seeing disc) rotated to the parallactic angle to avoid atmospheric slit
losses.  When using EMMI at the 3.5\,m ESO NTT, the spectral coverage required
for obtaining Str\"omgren $c_1$ coefficients from the spectra
($3200\,\mbox{\AA}<\lambda<4900\,\mbox{\AA}$) limits the maximum possible
dispersion to $1.8$\,{\AA} per pixel (grating \#4), since in the blue arm of
EMMI a 1\,k CCD is the only available choice. The pixel size is $0\farcs 37$,
so that at $\mbox{seeing}<1\farcs 2$, a spectral resolution of $<6$\,{\AA}
results.  Exposure times for obtaining $S/N>30$ at \ion{Ca}{2}~K are $5$\,min for stars
of $B<17.0$.  In the case where stars exhibit a very weak \ion{Ca}{2}~K line, as
recognized from online-reduced spectra, an additional, longer, exposure with
narrow ($1\farcs 0$) slit is obtained. The average total exposure time per
object is typically $10$\,min, which makes it possible to observe $\sim 30$
metal-poor candidates per night.

The spectra are shifted into the rest frame by cross-correlation with a model
spectrum of similar stellar parameters, and applying the appropriate radial
velocity correction. Note that the radial velocities derived are not useful
measurements in themselves, since the precise position of the object in the
(wide) slit is not known. Therefore, zero-point offsets in wavelength can
occur.

\begin{figure}[htbp]
  \begin{center}
    \leavevmode
    \epsfig{file=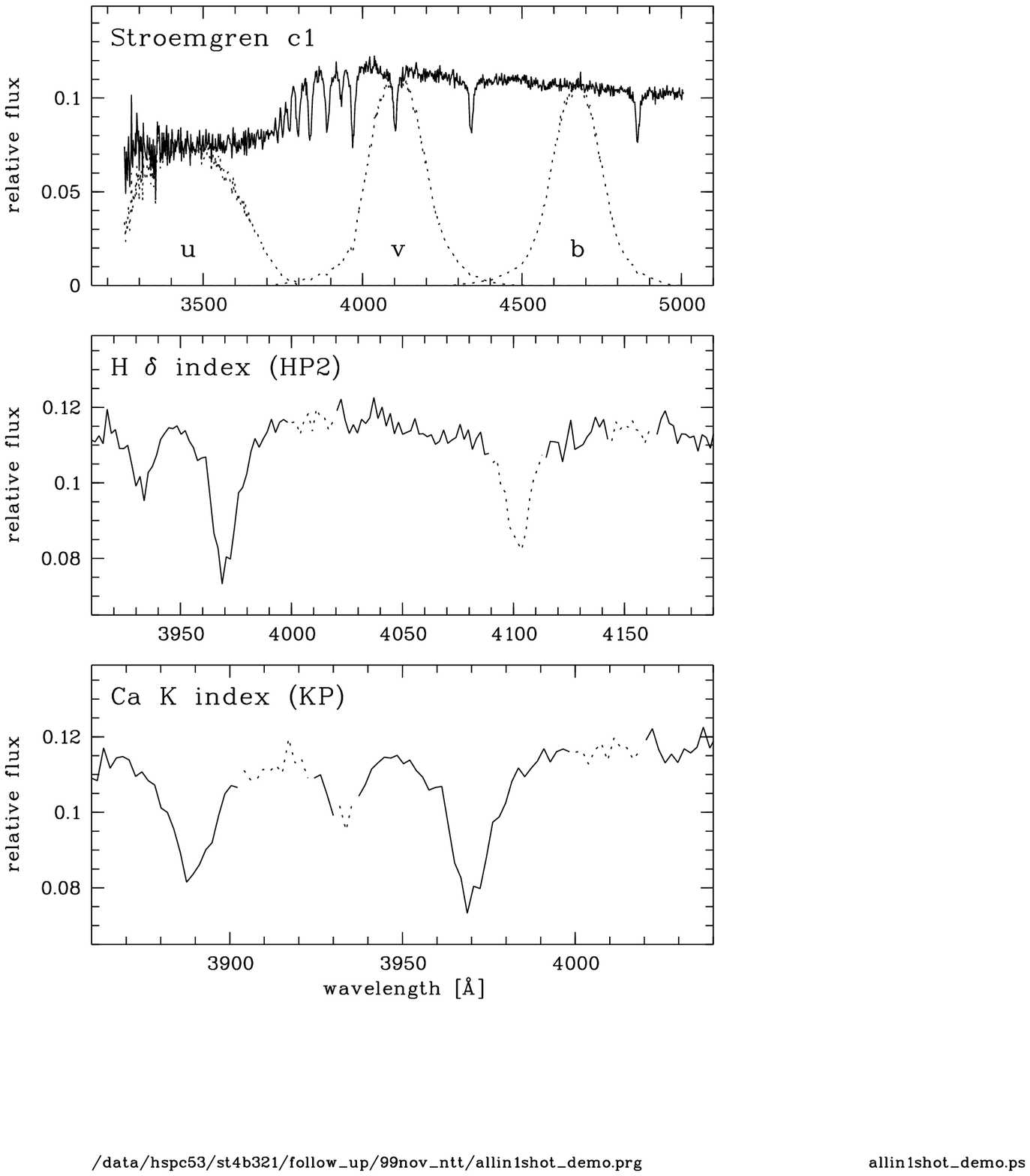, clip=, width=10cm,
      bbllx=69, bblly=240, bburx=400, bbury=697}
  \end{center}
  \caption{\label{allin1shot_demo} Determination of $c_1$, HP2, and KP from
    photometric, moderate resolution ($\sim 5$\,{\AA}) spectra obtained with
    EMMI attached to the 3.5\,m ESO NTT. Dashed lines in the lower two panels
    indicate continuum and line passbands used for the computation of HP2 and
    KP, respectively. }  
\end{figure}

Three features are used for determination of the stellar parameters [Fe/H],
$T_{\mbox{\scriptsize eff}}$, and $\log g$: the Str\"omgren-coefficient
$c_1=(u-b)-(v-b)$, the H$\delta$ index HP2, and the \ion{Ca}{2}~K index KP
\citep[for a definition see][]{HKrecalib}. The $c_1$ index is determined directly
from the spectra by multiplication with filter response curves and integration
over the appropriate wavelength range (see Fig. \ref{allin1shot_demo}). The
internal accuracy achieved is $\sigma_{c_1}=0.022$\,mag, which compares
favorably with errors from photoelectrically measured indices.

Stellar parameters are derived by using the following set of equations:
\begin{eqnarray}
  T_{\mbox{\scriptsize eff}} &=&
  a_{11}+a_{12}\cdot c_1+a_{13}\cdot\mbox{HP2}\label{TeffEq}\\ 
  \log g &=& a_{21}+a_{22}\cdot c_1+a_{23}\cdot\mbox{HP2}\label{loggEq}\\ 
  \left[\mbox{Fe/H}\right] &=& a_{31}+a_{32}\cdot\mbox{HP2}
  +a_{33}\cdot\mbox{KP}\label{FeHEq}
\end{eqnarray}
The coefficients $a_{ij}$ have been determined from least squares fits to a
dense grid of model spectra, defined by the following grid points:
\begin{eqnarray*}
  T_{\mbox{\scriptsize eff}} &=& 5600(200)6800\,\mbox{K}\\
  \log g &=& 2.2(0.8)4.6\\
  \left[\mbox{Fe/H}\right] &=& -0.9(0.3)-3.6
\end{eqnarray*}

Using equations (\ref{TeffEq})--(\ref{FeHEq}), it was possible to reproduce
the stellar parameters of the model spectrum grid with the following accuracy:
\begin{eqnarray*}
  \sigma_{T_{\mbox{\tiny eff}}} &=& 24\,\mbox{K}\\
  \sigma_{\log g} &=& 0.21\\ 
  \sigma_{\left[\mbox{\tiny Fe/H}\right]} &=& 0.16.
\end{eqnarray*}    
Note that these are {\em internal\/} errors for {\em noise-free\/} spectra.
Unfortunately, due to lack of an independent test sample, it is not yet
possible to estimate the {\em real\/} accuracy of this approach. However,
experience with spectrum synthesis has shown that at the spectral resolution
used in the HES follow-up, errors in $\log g$ and [Fe/H] are typically twice as
high as the numbers above, and errors in $T_{\mbox{\scriptsize eff}}$ are typically
$\sim 200$\,K.

\section{Discussion and conclusion}

We have compared two large, ongoing, surveys for metal-poor stars in the
Galactic halo, namely, the HK survey, and the HES.  Both surveys are in
the position to provide targets for observations with Subaru HDS {\em now}.
However, follow-up observations of HES stars have just been started, whereas
the HK survey has already produced a list of $\sim 4700$ stars with estimates
of [Fe/H] typically precise to $\pm 0.2$\,dex, on the order of 100 of which
exhibit the lowest abundances ever found for stars in the Galaxy.

Selection of metal-poor candidates at the main-sequence turnoff in the HES by
automatic classification is $\sim 3\times$/$\sim 7\times$ more efficient as
compared to visual inspection in the HK survey with/without pre-selection by
$BV$ photometry. This is very remarkable considering the fact that the
spectral resolution of the HES is $2\times$ {\em lower\/} than in the HK
survey. Reasons for the higher efficiency are the larger spectral coverage of
the HES, better quality of the HES spectra, and the automated, quantitative
selection, which is probably more precise than the selection by eye. Moreover,
we have intentionally observed class a and b candidates only, because we
wanted to explore what the {\em maximum possible\/} efficiency is. Simulations
we have carried out indicate that, in exchange for a high EY of truly
metal-poor stars, one has to sacrifice completeness of the candidate sample on
the order of 50\,\%. Thus, the EY of a selection aimed at compiling a {\em
  complete\/} sample of metal-poor stars by means of including class c
candidates and candidates from complementary selection methods (e.g. the
\ion{Ca}{2}~K index method), too, will be proportionately lower.

The follow-up technique used in the HK survey results in determinations of
[Fe/H] precise to $\pm 0.2$\,dex; the precision of the HES technique remains
to be evaluated. The advantage of the ``all in one shot''-technique used in
the HES is that no photometry is needed in addition to moderate resolution
spectra. However, a drawback is that no useful radial velocities can be measured
from spectra obtained with the wide slit, since the object position within the
slit is not precisely known, so that unknown zero-point offsets in wavelength
occur.  We are presently exploring the use of artificial neural network
methodology which might be able to recover the required stellar parameters with
sufficient accuracy from spectra taken with a narrow slit, so that radial
velocity information could be obtained simultaneously
\citep[see][]{Quetal:1998,Snideretal:2000}.

When compiling target lists for high-resolution observations, combining stars
from both surveys, it is important to take into account their different
abundance scales. An offset of $0.5$\,dex has to be subtracted from
[Fe/H] estimates obtained from the HES follow-up, when they are compared with
[Fe/H] values derived from the HK survey.  Since the limiting magnitude for
metal-poor stars in the HES is $\sim 17.0$, and ``saturated'' objects are
excluded from the selection procedure, the HES provides mainly fainter
candidates, in the magnitude range $14.0<B<17.0$, whereas the HK survey is able
to provide bright candidates in the range $11.0 < B <15.5$.

The HES is $\sim 1.5$\,mag deeper than the HK survey. Therefore, the former
can increase the total survey volume for metal-poor stars by a factor of 8,
taking into account common areas and the magnitude ranges of both surveys. We
estimate that the total number of stars at [Fe/H]$<-3.0$ known today, $\sim
100$, can be increased to $\sim 800$ by the HES, provided that follow-up
observations can be obtained for all candidates.  Extension of the procedures
described above for the inclusion of cooler stars could easily raise the
number of stars with [Fe/H] $< -3.0$ to 1000 or more.

Object lists formed from a combination of targets from both the HK and HES
surveys will be able to keep {\em all\/} $8$\,m class telescopes busy for at
least the next several years.  Observations of these stars are sure to provide
the astronomical community with many new insights (see Beers, this volume, for
an extensive listing).  However, a potential ``metal-poor star disaster'' is
only a few years away, in the sense that a much larger database of candidates
will be necessary to address the many {\em new\/} questions which are sure to
arise from the first-pass 8\,m-class telescope follow-ups
\citep[see][]{TimsFirstStarsComments}.  Now is the time to expand efforts to
obtain large numbers of new metal-deficient stars from follow-up of the HES
and HK survey candidates!

\acknowledgements

N.C. thanks the organizers of the Workshop for financial aid and Fujimoto-san
for private lessons in using Tokyo's public transport system. J. Reetz and T.
Gehren contributed to the search for metal-poor stars in the HES by providing
model atmospheres, SIU (a tool for spectrum analysis), repeated hospitality at
their institute, and many discussions. This work is supported by Deutsche
Forschungsgemeinschaft under grant Re~353/40--3.  T.C.B.  is grateful to his
many friends at NAO, and within the entire Japanese astronomical community,
for their long-term collaborations and discussions, and further acknowledges
partial support of this work from grant AST95-29454 from the National Science
Foundation.

\bibliography{classification,HES,mphs,ncastro,ncpublications,quasar}
\bibliographystyle{gagsm}


\end{document}